\documentclass[10pt,conference]{IEEEtran}
\IEEEoverridecommandlockouts
\usepackage{cite}
\usepackage{amsmath,amssymb,amsfonts}
\usepackage{algorithmic}
\usepackage{graphicx}
\usepackage{textcomp}
\usepackage{xcolor}
\def\BibTeX{{\rm B\kern-.05em{\sc i\kern-.025em b}\kern-.08em
    T\kern-.1667em\lower.7ex\hbox{E}\kern-.125emX}}

\usepackage{subcaption}
\usepackage{pbalance}

\newcommand{\argmin}{\mathop{\rm arg~min}\limits}

\makeatletter
\newcommand{\linebreakand}{% 
  \end{@IEEEauthorhalign}
  \hfill\mbox{}\par
  \mbox{}\hfill\begin{@IEEEauthorhalign}
}
\makeatother
\newcommand{\bm}[1]{{\mbox{\boldmath $#1$}}}
\begin{document}
 
\title{Evasive Ransomware Attacks Using Low-level Behavioral Adversarial Examples}
 
\author{%
\IEEEauthorblockN{Manabu Hirano}
\IEEEauthorblockA{\textit{Department of Information and Computer Engineering} \\
\textit{National Institute of Technology, Toyota College}\\
Toyota, Japan \\
hirano@toyota-ct.ac.jp}
\and
\IEEEauthorblockN{Ryotaro Kobayashi}
\IEEEauthorblockA{\textit{Faculty of Informatics} \\
\textit{Kogakuin University}\\
Tokyo, Japan \\
ryo.kobayashi@cc.kogakuin.ac.jp}
}

\maketitle

\begin{abstract}
  Protecting state-of-the-art AI-based cybersecurity defense systems from cyber attacks is crucial. Attackers create adversarial examples by adding small changes (i.e., perturbations) to the attack features to evade or fool the deep learning model. This paper introduces the concept of low-level behavioral adversarial examples and its threat model of evasive ransomware. We formulate the method and the threat model to generate the optimal source code of evasive malware. We then examine the method using the leaked source code of Conti ransomware with the micro-behavior control function. The micro-behavior control function is our test component to simulate changing source code in ransomware;  ransomware's behavior can be changed by specifying the number of threads, file encryption ratio, and delay after file encryption at the boot time. We evaluated how much an attacker can control the behavioral features of ransomware using the micro-behavior control function to decrease the detection rate of a ransomware detector.
\end{abstract}


%%%%%%%%%%%%%%%%%%%%%%%%%%
%% Introduction
%%%%%%%%%%%%%%%%%%%%%%%%%%
\section{Introduction}\label{sec:intro}

%%% AI in cybersecurity and adversarial examples
The term \emph{adversarial example} was first introduced by Szegedy et al. \cite{szegedy2013intriguing}; they found that applying an imperceptible non-random perturbation to a test image (i.e., the image is called \emph{adversarial example}) can arbitrarily
change the neural network’s prediction. As the application of deep learning in cybersecurity increased, the attacks against AI-based defense systems increased. Macas et al. presented a survey of attacks and defenses in deep-learning-based cybersecurity systems from the perspective of adversarial examples \cite{macas2024adversarial}; the system includes malware detection, botnet detection, network intrusion detection, fraud detection, and cyber-physical system (CPS) security. In malware detection, they described attacks using adversarial examples of static features obtained from a Portable Executable (PE) file or Android Application Package (APK), an image created from malware binary, and Control Flow Graph (CFG) features. An adversarial example is created using methods such as gradient-based attacks, optimization-based attacks, and generative adversarial network (GAN)-based attacks. This paper focuses on ransomware detection using ransomware’s behavioral features.

Ransomware attacks can be detected using indicators obtained from Application Programming Interface (API) and system call monitoring, Input and Output (I/O) monitoring, file system operation monitoring, etc\cite{begovic2023cryptographic}. Machine-learning-based ransomware detection method uses structural features, behavioral features, and both structural and behavioral features \cite{oz2022survey}; the behavioral features include hardware behavior, file system behavior, network traffic behavior, API call behavior, and a set of all the behavioral features. This paper employs low-level storage and memory access patterns as behavioral features. We first describe the difficulty of creating evasive ransomware against behavioral-based ransomware detectors.

%%% Problem Statement
\subsection{Difficulty on generating evasive malware from behavioral adversarial examples}

\begin{figure*}[tb]
\centering
\includegraphics[scale=0.65]{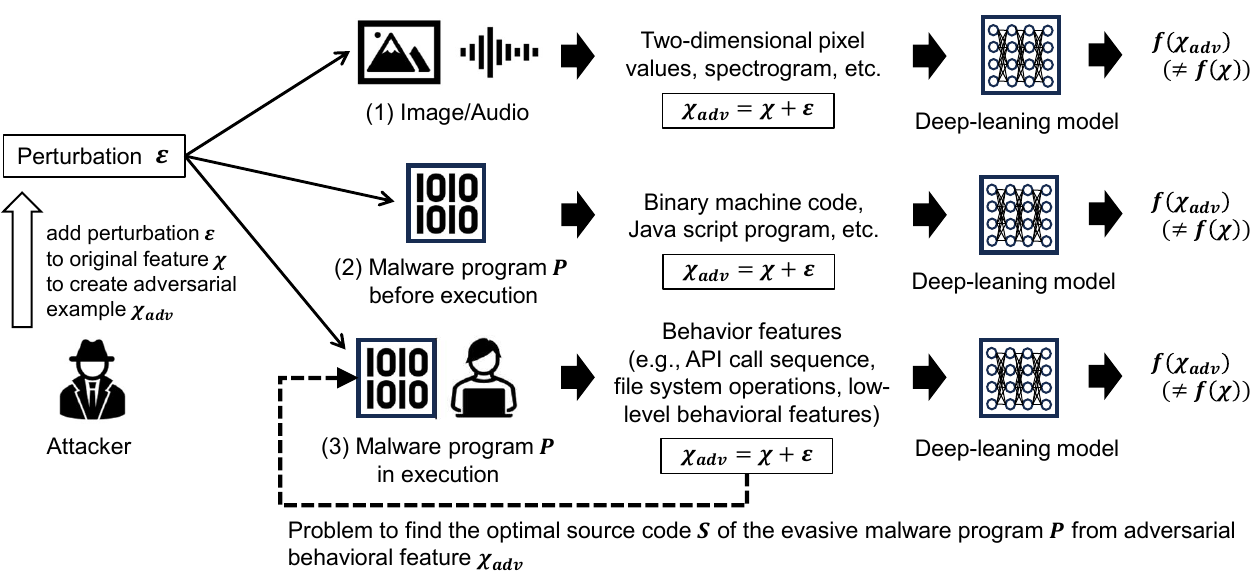}
\caption{Evasion attack scenarios using adversarial examples $\bm{\chi}_{adv}$ on image/audio classifier, static-analysis-based malware detector, and dynamic-analysis-based malware detector.}
\label{fig:adversarial-attack}
\end{figure*}

Fig.~\ref{fig:adversarial-attack} shows evasion attack scenarios using adversarial examples $\bm{\chi}_{adv}$ on image or audio classifier, static-analysis-based malware detector, and dynamic-analysis-based malware detector.
The first example is an image or audio classifier using a deep-learning model $f$. An attacker aims to cause the deep-learning model to return an incorrect class. The attacker adds subtle perturbation (e.g., slight noise) $\bm{\epsilon}$ that is not recognized by human perception to original image pixel values or audio spectrogram $\bm{\chi}$ to create an \emph{adversarial example} $\bm{\chi}_{adv}$ (= $\bm{\chi}$ + $\bm{\epsilon}$). A deep-learning model $f(\bm{\chi})$ returns a class label for the input $\bm{\chi}$. 
Evasion attacks succeed when an attacker inputs \emph{adversarial example} $\bm{\chi}_{adv}$ to the target deep-learning model $f$ and obtains incorrect class label $f(\bm{\chi}_{adv})$ not equal to the correct class label $f(\bm{\chi})$. %% Creating adversarial examples of images and audio is relatively easy since attackers can directly insert perturbations to original examples.

The second example is a malware detector using a deep-learning model $f$. An attacker aims to evade the malware detector by causing the deep-learning model to return an incorrect class (i.e., benign class). The attacker adds a perturbation $\bm{\epsilon}$ to the malware program $\mathcal{P}$ to evade the target malware detector (i.e., deep-learning model) $f$. Evasion attack succeeds when the attacker inputs \emph{adversarial example} $\bm{\chi}_{adv}$ to the target deep-learning model $f$ and obtained incorrect class $f(\bm{\chi}_{adv})$ not equals to $f(\bm{\chi})$. 
Hu and Tan presented an evasion attack method named \emph{MalGAN} \cite{malgan}; they assume the target deep-learning model $f$ uses a 160-dimensional binary vector $\bm{\chi}$ to represent the malware program's static feature. The 160-dimensional binary vector $\bm{\chi}$ consists of a series of binary values $\{0,1\}$ that represents the existence of 160 representative API calls (i.e., $1$ means the API call exists in the malware program, $0$ means it does not exist). Since deletion of critical API calls affects functions of malware (e.g., deleting \emph{WriteFile} API call affects ransomware's destruction function), \emph{MalGAN} only adds perturbation $\bm{\epsilon}$ to API calls that do not affect the original malware functions. The attacker can create a malware program by adding some binary code that calls the API calls (i.e., the code corresponding to the perturbation $\bm{\epsilon}$) at the places that do not affect the original function (e.g., slack space) in the original binary program $\mathcal{P}$. Creating an evasive malware program $\mathcal{P}$ against static-analysis-based malware detector using an \emph{adversarial example} $\bm{\chi_{adv}}$ is relatively easy when attackers can directly insert binary code (i.e., perturbation $\bm{\epsilon}$) to the original binary program $\mathcal{P}$ without corrupting its functions.

The last example is a behavioral-based malware detector using a deep-learning model $f$. An attacker's goal is to evade the malware detector by causing the deep-learning model to return an incorrect class (i.e., benign class); this time, the malware detector (i.e., deep-learning model) $f$ uses a behavioral feature $\bm{\chi}$ obtained when the malware is executed. The attackers added perturbation $\bm{\epsilon}$ to the original behavioral feature $\bm{\chi}$ to create an \emph{behavioral adversarial example} $\bm{\chi_{adv}}$ (= $\bm{\chi}$ + $\bm{\epsilon}$). Evasion attack succeeds when the attacker inputs $\bm{\chi}_{adv}$ to the target deep-learning model $f$ and obtained incorrect class $f(\bm{\chi}_{adv})$ not equals to $f(\bm{\chi})$. The attacker's goal is to find the optimal source code $\mathcal{S}_{adv}$ of the evasive malware program $\mathcal{P}_{adv}$ from \emph{behavioral adversarial examples} $\bm{\chi_{adv}}$.

\subsection{Contribution and organization of this paper}

Although creating \emph{behavioral adversarial examples} $\bm{\chi_{adv}}$ at the feature level is relatively easy, creating an evasive malware program (i.e., source code) that produces the \emph{behavioral adversarial examples} $\bm{\chi_{adv}}$ is challenging as shown in the example (3) of Fig.~\ref{fig:adversarial-attack}. This paper's contributions are as follows:

\begin{itemize}
\item We formulate a method and threat model in creating evasive ransomware program; an attacker finds the optimal source code $\mathcal{S}$ of evasive ransomware program $\mathcal{P}$ that produces \emph{behavioral adversarial examples} $\bm{\chi_{adv}}$ to evade the target behavioral-based ransomware detector $f$.
\item We present the \emph{micro-behavior control function} to examine the formulated method in a real-world ransomware sample using the leaked Conti ransomware's source code.
\item We evaluated how much an attacker can change behavioral features $\bm{\chi_{adv}}$ by modifying the ransomware source code. At last, we evaluated the success rate of evasive ransomware attacks.
\end{itemize}
  
The rest of this paper is organized as follows. Section \ref{sec:formulation} presents the formulation of the evasive ransomware problem using \emph{behavioral adversarial examples}. Section \ref{sec:feature-description} describes the low-level behavioral features used in this paper. Section \ref{sec:micro-behavior} presents the design and implementation of the \emph{micro-behavior control function} to examine the formulated method. Section \ref{sec:evaluation} shows how much the \emph{micro-behavior control function} can change the behavioral features; section \ref{sec:success-rate} shows how much the \emph{micro-behavior control function} can evade the ransomware detector. Section \ref{sec:discussion} describes the limitations and future work. We conclude the summary in Section \ref{sec:conclusion}.

%%%%%%%%%%%%%%%%%%%%%%%%%%%%%%%%%%%%%%%%%%%%%%%%%%%%%%%%%%%%%%%%%
%% Threat model
%%%%%%%%%%%%%%%%%%%%%%%%%%%%%%%%%%%%%%%%%%%%%%%%%%%%%%%%%%%%%%%%%
\section{Formulation of behavioral adversarial examples and threat model on evasive ransomware problem}\label{sec:formulation}

We aim to examine how the attackers can change a behavioral feature $\bm{\chi}$ of ransomware to evade the target behavioral-based ransomware detector represented as a deep-learning model $f$ by adding small perturbations $\bm{\epsilon}$. In this paper, we define \emph{behavioral adversarial examples} $\bm{\chi_{adv}}$ of ransomware and threat model as follows. 

\begin{itemize}
\item Attackers can create an approximation of the target ransomware detector (i.e., \emph{substitute} model $f$) by submitting sufficient queries and receiving responses to learn the input and output mappings. We assume that attackers know what behavioral features the target ransomware detector's deep learning model uses (i.e., \emph{grey box} attack). When the \emph{substitute} model $f$ became an imitation close enough to the target ransomware detector's deep learning model $f_{orig}$, attackers can use the \emph{substitute} model $f$ to create adversarial behavioral feature $\bm{\chi_{adv}}$ by submitting the unlimited number of queries and of responses.
\item Attackers' first goal is to find an adversarial behavioral feature $\bm{\chi_{adv}}$ that is classified as \emph{benign} by the target detector with minimum perturbation $\bm{\epsilon}$.
\item Attackers have access to a source code $\mathcal{S}_{orig}$ of ransomware; Attackers' final goal is to find the optimal source code $\mathcal{S}_{adv}$ that produces a behavioral adversarial feature $\bm{\chi_{adv}}$. The optimization (i.e., selecting the optimal source code from candidates) is performed by minimizing the ransomware's functional degradation $\mathcal{F}$ and performance degradation $\mathcal{P}$.

\end{itemize}

In our threat model, attackers can find adversarial behavioral feature $\bm{\chi_{adv}}$ in \eqref{eq:adversarial}. We assume the attacker aims to evade a binary classification model (i.e., ransomware or benign class).

\begin{equation}\label{eq:adversarial}
\bm{\chi_{adv}}\,=\,\bm{\chi} + \argmin_{\bm{\epsilon}} \Big\{||\bm{\epsilon}||_{p} : f(\bm{\chi_{adv}}) \ne f(\bm{\chi}) \Big\}
\end{equation}

\noindent where $\bm{\epsilon}$ is a tensor representing the minimum perturbation to the original behavioral feature $\bm{\chi}$ (i.e., $\bm{\chi_{adv}}\,=\,\bm{\chi} + \bm{\epsilon}$). $||\bm{\epsilon}||_{p}$ is an $L^{p}$-norm; it is Euclidean norm when $p$ is 2. A function $f(\bm{\chi})$ returns a class label (i.e., ransomware or benign); attackers modify original ransomware that produces a behavioral feature $\bm{\chi}$ to evade the target detector; the modified version of ransomware produces a behavioral adversarial feature $\bm{\chi_{adv}}$. When $f(\bm{\chi_{adv}}) \ne f(\bm{\chi})$ is satisfied, the attacker succeeds in evading the target detector using the modified version of ransomware that is misclassified in benign class; the perturbation $\bm{\epsilon}$ should be small as possible.

%%% Difficulty mapping epsilon and source code
However, unlike conventional adversarial examples problems such as image and audio perturbation, we need to generate the evasive ransomware program $\mathcal{P}_{adv}$ that can be executed on a computer. We need to formulate how to find the optimal source code $\mathcal{S}_{adv}$ of the evasive ransomware program $\mathcal{P}_{adv}$. We first formulate the mapping between a \emph{behavioral adversarial perturbation} $\bm{\epsilon}$ and source code $\mathcal{S}$ in \eqref{eq:mapping}.

\begin{equation}\label{eq:mapping}
\bm{\epsilon}_{i,j}\,=\,\bm{\mathcal{E}}(\mathcal{S}_i,\mathcal{V}_j)
\end{equation}

\noindent where $\bm{\mathcal{E}}(\mathcal{S}_i,\mathcal{V}_j)$ is a projection function from a source code candidate $\mathcal{S}_i$ and an execution environment $\mathcal{V}_j$ to an adversarial perturbation $\bm{\epsilon}$. For example, an evasive ransomware program $\mathcal{P}_{i,j}$ created from a source code $\mathcal{S}_i$ produces a behavioral adversarial feature $\bm{\chi}_{adv}$ (=$\bm{\chi}+\bm{\epsilon}_{i,j}$) when it is executed on an environment $\mathcal{V}_j$ (e.g., specifications of computers and network including the number of Central Processing Unit (CPU) cores and network throughput). Although evasive ransomware should have some adaptive mechanism to reproduce behavioral adversarial examples on various computer and network environments, we tested only one computer and network environment (i.e., constant $\mathcal{V}$) in this paper. The advanced adaptive mechanism of evasive ransomware that can be executed on various environments is possible future work. 

%%% The optimal source code selection from candidates
The optimal source code selection is performed in \eqref{eq:source}.

\begin{equation}\label{eq:source}
\begin{array}{r}
\mathcal{S}_{adv}\,=\,\argmin_{\mathcal{S}} \Big\{
 \alpha\mathcal{F}(\mathcal{S})+\beta\mathcal{P}(\mathcal{S},\mathcal{V}) : \\
 f(\bm{\chi}+\bm{\mathcal{E}}(\mathcal{S},\mathcal{V})) \ne f(\bm{\chi})
\Big\}
\end{array}
\end{equation}

\noindent where function $\mathcal{F}(\mathcal{S})$ returns the degradation in the number of functions original ransomware has when attackers introduce perturbation $\bm{\epsilon}$; the functions include file enumeration and encryption, network drive encryption, data exfiltration, lateral movement, privilege escalation, Command and Control communication. On the other hand, the function $\mathcal{P}(\mathcal{S},\mathcal{V})$ returns the performance degradation of ransomware (e.g., the number of encrypted files per second, the throughput of data exfiltration) on the execution environment $\mathcal{V}$. Attackers balance functional degradation $\mathcal{F}$ and performance degradation $\mathcal{P}$ by deciding appropriate coefficients $\alpha$ and $\beta$ in \eqref{eq:source} to select the optimal source code from candidates while the minimum condition for a successful evasion attack \eqref{eq:adversarial} is satisfied.

%%% Bridge to the next section
The contribution of this paper is to examine a method to find an optimal source code $\mathcal{S}_{adv}$ of ransomware that produces a behavioral adversarial feature $\bm{\chi_{adv}}$ that meets above requirements. The following section presents the detail of the behavioral feature $\bm{\chi}$ we used in this paper.

%%%%%%%%%%%%%%%%%%%%%%%%%%%%%%%%%%%%%%%%%%%%%%%%%%%%%%%%%%%%%%%%%
%% Low-level Behavioral features used in ransomware detector
%%%%%%%%%%%%%%%%%%%%%%%%%%%%%%%%%%%%%%%%%%%%%%%%%%%%%%%%%%%%%%%%%
\section{Low-level behavioral features used in ransomware detector}
\label{sec:feature-description}

\begin{figure}[tb]
\centering
\includegraphics[scale=0.7]{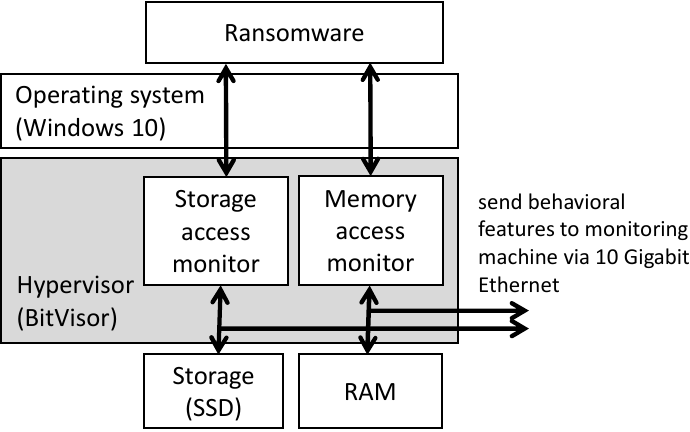}
\caption{Hypervisor-based monitoring system to collect low-level behavioral features \cite{HIRANO2022301314,HIRANO2025104202}.}
\label{fig:monitoring-system}
\end{figure}

%% How to obtain low-level behavioral features
We presented a hypervisor-based monitoring system to collect low-level behavioral features for detecting ransomware's destruction phase \cite{HIRANO2022301314, HIRANO2025104202}. In our defense-in-depth approach, we employed the thin hypervisor as an additional protection layer to the conventional OS-level protection layer. Fig.~\ref{fig:monitoring-system} shows the hypervisor-based monitoring system; the low-level behavioral features collected in the hypervisor layer consist of access patterns on storage devices (e.g., Solid State Drive) and memory access patterns on Random Access Memory (RAM). We developed the monitoring functions using a thin hypervisor named BitVisor \cite{bitvisor}. The storage access patterns are obtained using an extended Advanced Host Controller Interface (AHCI) para-pass through driver \cite{HIRANO2022301314}, while memory access patterns are obtained using a hardware-assisted memory virtualization technology \cite{HIRANO2025104202}, Intel's Extended Page Table (EPT) \cite{intel-sdm}; we examined a deep-learning-based ransomware detector trained using the low-level behavioral features. Please refer to our paper \cite{HIRANO2025104202} for details on the hypervisor-based monitoring system.

\begin{figure*}[tb]
\centering
\includegraphics[scale=0.42]{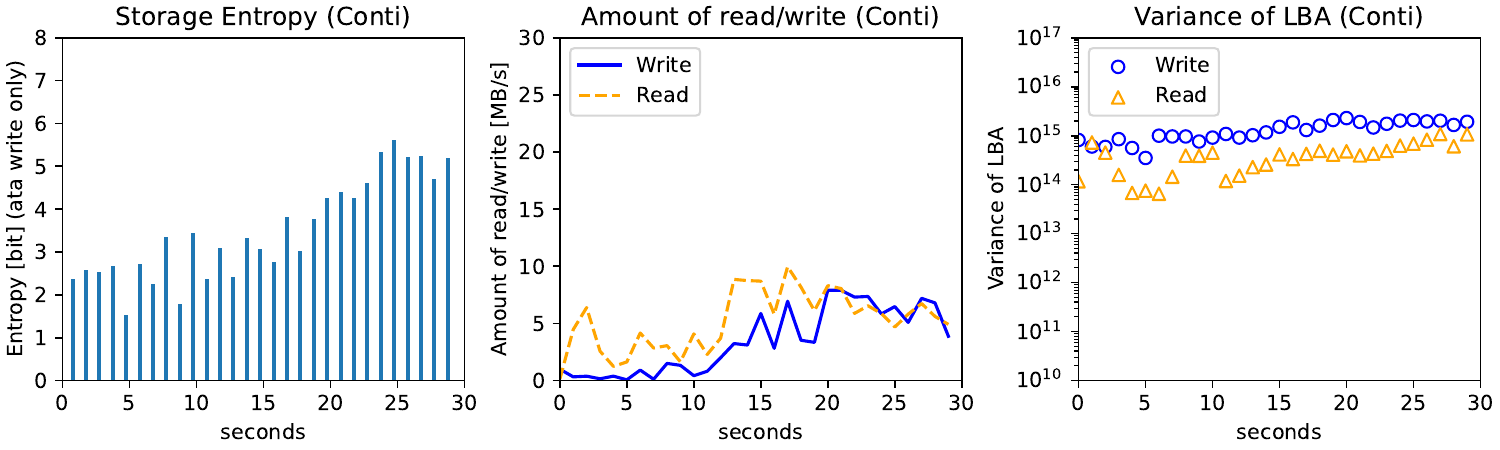}
\includegraphics[scale=0.42]{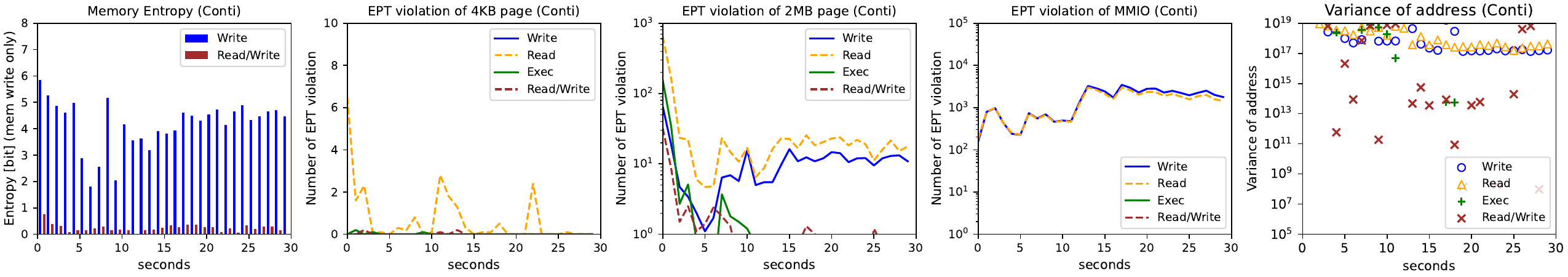}
\caption{Example of a storage feature $\bm{\chi_s}$ (top) and a memory feature $\bm{\chi_m}$ (bottom) of Conti ransomware sample in the RanSMAP dataset, open behavioral feature dataset of ransomware storage and memory access patterns\cite{HIRANO2025104202,RanSMAP-repo}. The graphs were created using average behavioral features of 10 trials on the computer with Intel Core i3 and DDR4-2133 16GB RAM.}
\label{fig:Conti}
\end{figure*}

%% What is the behavioral features: definition of the symbols
Fig.~\ref{fig:Conti} shows graphs of a behavioral feature $\bm{\chi}$ obtained using the developed hypervisor-based monitoring system; a low-level behavioral feature $\bm{\chi}$ consists of a storage feature $\bm{\chi_s}$ and a memory feature $\bm{\chi_m}$. A storage feature $\bm{\chi_s}$ consists of the following five-dimensional feature vectors ($\bm{x_0}$ to $\bm{x_4}$): entropy of written blocks (top-left graph), write and read throughput in Byte/s (center-top graph), and variance of written and read Logical Block Address (LBA) on storage devices (top-right graph). On the other hand, a memory feature $\bm{\chi_m}$ consists of the following 18-dimensional feature vectors ($\bm{x_5}$ to $\bm{x_{22}}$): entropy in write and read-write operations (bottom-leftmost graph), the number of EPT violations (i.e., the number of memory accesses) for 4KiB, 2MiB, and Memory Mapped Input Output accesses of write, read, instruction fetch, and read-write operations (the three bottom-center graphs), and variance of accessed Guest Physical Address (GPA) of write, read, instruction fetch, and read-write operations (bottom-rightmost graph). 
%% Why we use EPT violation
We counted the number of memory accesses using EPT violation. An EPT violation occurs when a hypervisor does not have the accessed GPA page table entry in the Extended Page Table (EPT). Once the address is translated, the EPT violation never occurs at the same address. Therefore, our system periodically flushes a Translation Lookaside Buffer (TLB) to cause intentional EPT violations. Performance degradation caused by the monitoring system was examined in our previous paper \cite{HIRANO-CSR2022}, which presented the use of memory access patterns to detect ransomware. We presented the details of the feature engineering in another paper \cite{HIRANO2025104202}, which showed the analysis of our dataset that contains memory and storage access patterns of ransomware.

\begin{figure}[tb]
\centering
\includegraphics[scale=0.42]{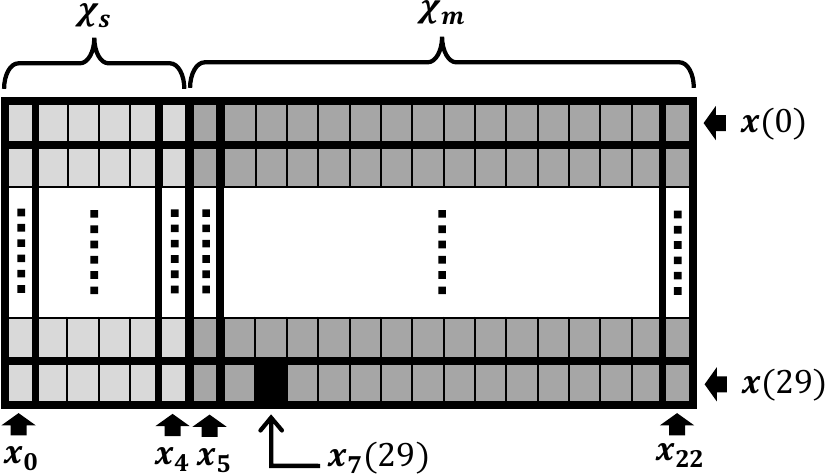}
\caption{Structure of a behavioral feature $\bm{\chi}$ consisting of $\bm{\chi_s}$ and $\bm{\chi_m}$ in 30 s after executing a ransomware sample or benign application ($T_{window}$\,=\,1 s and $T_d$\,=\,30 s).}
\label{fig:feature-description}
\end{figure}

Fig.~\ref{fig:feature-description} shows a structure of a behavioral feature $\bm{\chi}$ that consists of a storage feature $\bm{\chi_s}$ and a memory feature $\bm{\chi_m}$; they are collected in 30 s after executing a ransomware sample (i.e., Conti, Darkside, LockBit, REvil, Ryuk, and WannaCry) or a benign application (i.e., AESCrypt, Firefox, Idle, Office, SDelete, and Zip). The graphs is created using $T_{window}$\,=\,1 s and $T_{d}$\,=\,30; $T_{window}$ is a period to calculate a feature vector $\bm{x}$. $T_d$ is the duration of access patterns used in detecting ransomware; therefore, $T_{d}$ is also referred to as the detection time of ransomware. As we can see in Fig.~\ref{fig:feature-description}, $\bm{\chi}\,=\,\{\bm{\chi_s}, \bm{\chi_m}\}\,=\,\{\bm{x}(0), \bm{x}(1), \bm{x}(2),..., \bm{x}(29)\}$. Each row represents a feature vector at a specific period; for example, $\bm{x}(0)$ is a 23-dimensional feature vector calculated using storage and memory access patterns between 0 s and 1 s. On the other hand, a behavioral feature $\bm{\chi}$ can also be expressed in $\bm{\chi}\,=\,\{\bm{\chi_s}, \bm{\chi_m}\}\,=\,\{ \bm{x_0}, \bm{x_1}, \bm{x_2},..., \bm{x_{22}}\}$. Each column represents a specific feature vector between 0 s and 29 s. For example, $\bm{x_0}\,=\,\{ \bm{x_0}(0), \bm{x_0}(1), \bm{x_0}(2),..., \bm{x_0}(29) \}$ is a feature vector of storage entropy between 0 s and 30 s. $\bm{x_i}(j)$ is an $i$th feature vector calculated using access patterns between $j$ s and $(j+1)$ s.
%%
%% Bridge to the next section
The following section describes how we can change behavioral feature $\bm{\chi}$ by changing the micro-behaviors of ransomware.

%%%%%%%%%%%%%%%%%%%%%%%%%%%%%%%%%%%%%%%%%%%%%%%%%%%%%%%%%%%%%%%%%
%% Micro-behavior control function of Conti ransomware
%%%%%%%%%%%%%%%%%%%%%%%%%%%%%%%%%%%%%%%%%%%%%%%%%%%%%%%%%%%%%%%%%
\section{Micro-behavior control function of evasive ransomware}
\label{sec:micro-behavior}

%% Why we created a micro-behavior control function
In Section \ref{sec:intro}, we described difficulty on generating an evasive malware program $\mathcal{P}$ (i.e., source code $\mathcal{S}$ corresponding to $\mathcal{P}$) that produces \emph{adversarial behavioral feature} $\bm{\chi}_{adv}$. In Section \ref{sec:formulation}, we formulated the optimal source code selection problem in \eqref{eq:source}. The goal of the problem in \eqref{eq:source} is to find the best source code $\mathcal{S}_{adv}$ from candidate source codes $\bm{\mathcal{S}} =\{\mathcal{S}_0, \mathcal{S}_1, \mathcal{S}_2, ... , \mathcal{S}_\infty\}$. $\mathcal{S}_i$ (= $\mathcal{S}_{orig}$ + $\mathcal{S}_{\epsilon_i}$) is a source code to produce a behavioral adversarial feature $\bm{\chi}_i$; $\mathcal{S}_{orig}$ is an original source code of ransomware and $\mathcal{S}_{\epsilon_i}$ is a patch to create $\mathcal{S}_{i}$ from $\mathcal{S}_{orig}$. Since the search space is vast and we do not have generative AI that outputs the optimal source code yet, we first test the method in limited source code space; in this paper, we simulated 24 patterns of source code candidates $\bm{\mathcal{S}} =\{\mathcal{S}_0, \mathcal{S}_1, \mathcal{S}_2, ..., \mathcal{S}_{23}\}$ using boot options of a modified version of Conti ransomware. The boot options consist of the number of threads, encryption ratio, and delay after file encryption; these parameters are used in our \emph{micro-behavior control function} of evasive ransomware. The micro-behavior control function tested in this paper produces 24 patterns of a behavioral feature $\bm{\chi}_i$. We examined the changes in the behavioral features and tested the success rates of evasion attacks on the target ransomware detector.

\begin{figure}[tb]
\centering
\includegraphics[scale=0.50]{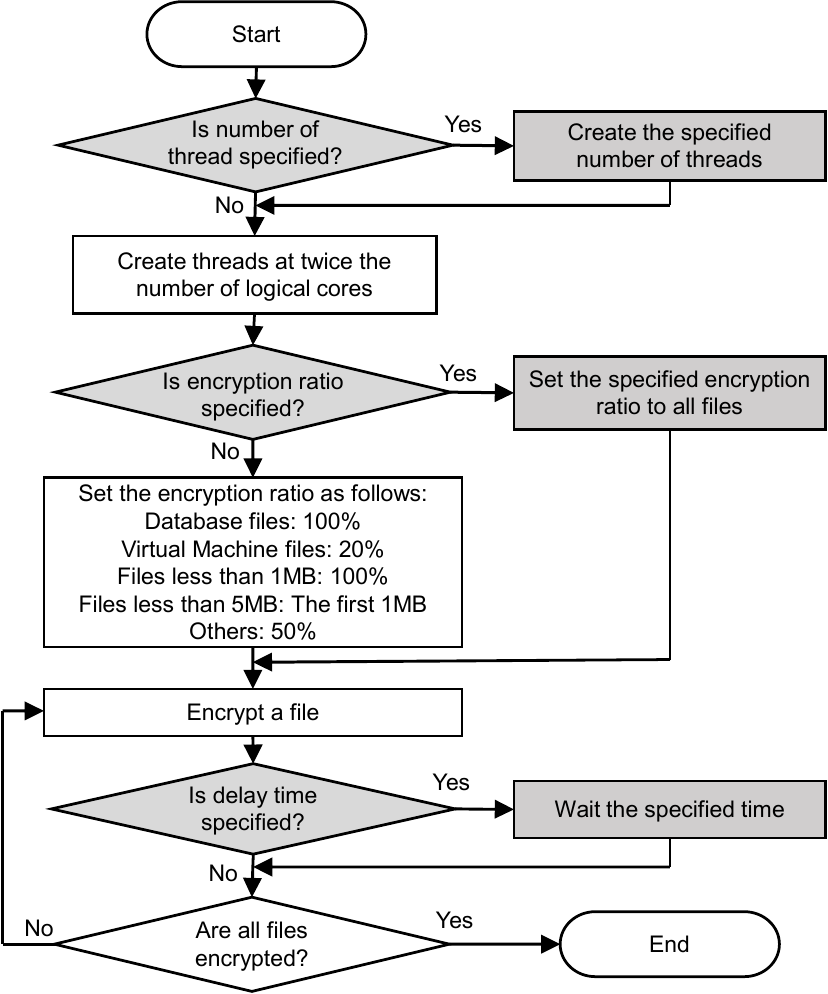}
\caption{Flow chart of the micro-behavior control function of Conti ransomware. The extended parts are shown in grey.}
\label{fig:flowchart}
\end{figure}

%% What is a micro-behavior of ransomware
Fig.~\ref{fig:flowchart} shows a flow chart of the micro-behavior control function of modified Conti ransomware. The extended parts of Conti ransomware are shown in grey. We control the micro-behaviors of Conti ransomware by changing the following three parameters that are specified at the execution time of ransomware: (1) the number of threads, (2) the encryption ratio per file, and (3) the delay after encrypting a file. We added the source code of the extended parts to the leaked source code of Conti ransomware \cite{leaked-conti}. The development and test were performed using the isolated lab environment. 

\begin{figure}[tb]
\centering
\includegraphics[scale=0.5]{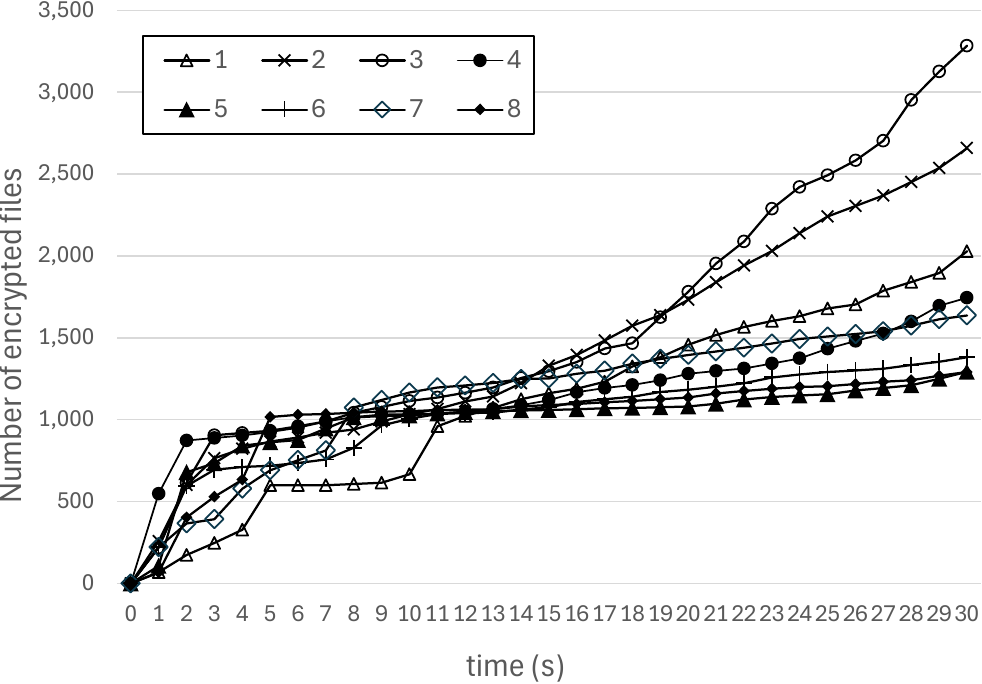}
\caption{The number of encrypted files in 30 s when the micro-behavior control function of Conti ransomware was executed using the number of threads between 1 and 8.}
\label{fig:thread-graph}
\end{figure}

%%%% Description of parameters
Table~\ref{tbl:spec} shows the computer's specification for developing and testing the micro-behavior control function of Conti ransomware. 
%% The first test of the micro-behavior control function
Fig.~\ref{fig:thread-graph} shows the changes in the number of encrypted files when we executed the Conti ransomware with the micro-behavior control function; we tested the number of threads between 1 and 8. Although we used the computer with eight logical cores (i.e., eight threads), the number of encrypted files decreased when we specified more than three threads. The original Conti ransomware uses twice the number of logical cores as the number of threads; therefore, the number of threads will be 16 when we execute the original Conti ransomware on the computer with a CPU that supports eight logical cores. The performance degradation in more than three threads on the test machine indicates that some ransomware samples are not correctly optimized for computers of various performances.
%%%
This study aims to change the micro-behavior of ransomware; we use the number of threads between 1 and 3 in the later section since the number of threads and the number of encrypted files are proportional in the range between 1 and 3 on the test machine. Please note that the behavior depends on the computer on which ransomware is executed (i.e., an environment $\mathcal{V}$); for example, the range will change if we use high-performance file servers with high-grade CPU and storage. In this paper, we tested our micro-behavior control function only on the test machines shown in Table~\ref{tbl:spec}. Examining the micro-behavior control function of the computers of various performances is possible for future work.

\begin{table}[t]
\caption{Specification of the computer used in developing the micro-behavior control function of evasive ransomware.}
\label{tbl:spec}
\begin{center}
\begin{tabular}{lp{60mm}}
\hline
\noalign{\vskip 1.0pt}
CPU & Intel Core i3 12100 \\
 & 4 P-cores, 8 threads \\
RAM & DDR4 2133 8GiB x 2 \\
Motherboard & ASRock B660M-HDV \\
Solid State Drive & Crucial CT240BX, CT250MX, Samsung 840 \\
Hypervisor & BitVisor downloaded on 25th Jan. 2022 (8c129a1) \\
\noalign{\vskip 0.5pt}
\hline % last hline
\end{tabular}
\end{center}
\end{table}

%%%%%%%%%%%%%%%%%%%%%%%%%%%%%%%%%%%%%%%%%%%%%%%%%%%%
%% Evaluation of micro-behavior control function
%%%%%%%%%%%%%%%%%%%%%%%%%%%%%%%%%%%%%%%%%%%%%%%%%%%%
\section{Evaluation of micro-behavior contorol function of evasive ransomware}\label{sec:evaluation}

\begin{figure*}[tb]
\centering
\includegraphics[scale=0.7]{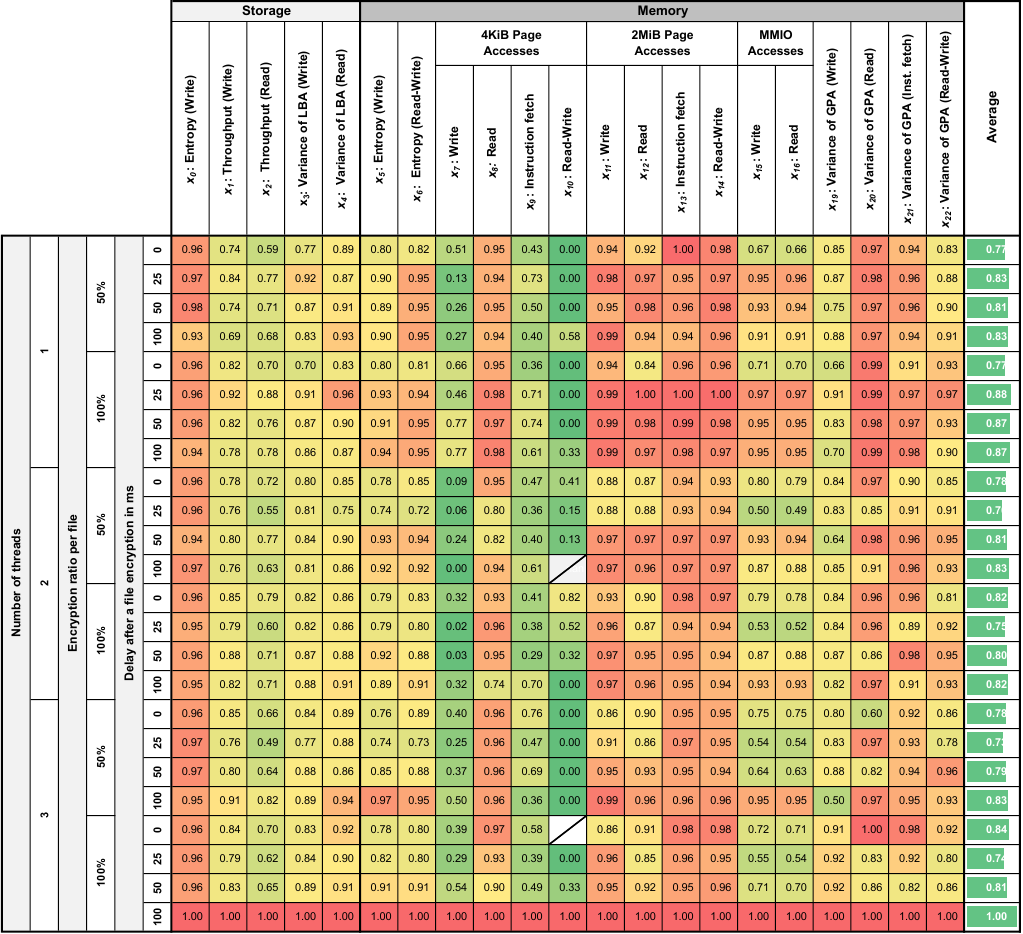}
\caption{Cosine similarity between adversarial behavioral features $\chi_{adv}$ of modified Conti ransomware with the micro-behavior control function. Cosine similarity is calculated using the parameters of $\mathit{P_{thread}}$\,=\,$3$, $\mathit{P_{ratio}}$\,=\,$100\%$, and $\mathit{P_{delay}}$\,=\,$100$ ms.}
\label{fig:similarity-mod-mod}
\end{figure*}

How much can a ransomware author control behavioral feature $\bm{\chi}$ by changing the ransomware's source code? This section presents an analysis of changes in the behavioral feature $\bm{\chi}$ obtained from the modified Conti ransomware with the micro-behavior control function that simulates changes in source code.
%%%%
Fig.~\ref{fig:similarity-mod-mod} shows cosine similarity between adversarial behavioral features $\chi_{adv}$ of modified Conti ransomware with the micro-behavior control function. The tested parameters are $\mathit{P_{thread}}$\,=\,\{1, 2, 3\}, $\mathit{P_{ratio}}$\,=\,\{50\%, 100\%\}, and $\mathit{P_{delay}}$\,=\,\{0 ms, 25 ms, 50 ms, 100 ms\}. The number of combinations of parameter $P$ is 24 in total. Please note that we excluded $\bm{x}_{17}$ and $\bm{x}_{18}$ since there were no MMIO accesses of instruction fetch and read-write.
we calculated cosine similarity between $\bm{x_{(i,base)}}$ and $\bm{x_{(i,k)}}$ in \eqref{similarity-mod-mod}.

\begin{equation}\label{similarity-mod-mod}
Sim(\bm{x_{(i,base)}},\bm{x_{(i,k)}}) 
= \dfrac{\bm{x_{(i,base)}}\,\bm{\cdot}\,\bm{x_{(i,k)}}}{||\bm{x_{(i,base)}}||\,||\bm{x_{(i,k)}}||}
\end{equation}

\noindent where $\bm{x_{(i,base)}}$ is an $i$th behavioral feature vector $\bm{x_i}$ of the modified ransomware using the parameter set $base$ that uses $\mathit{P_{thread}}$\,=\,$3$, $\mathit{P_{ratio}}$\,=\,$100\%$, and $\mathit{P_{delay}}$\,=\,$100$ ms; we used the bottom row of Fig.~\ref{fig:similarity-mod-mod} as baseline to calculate all other cosine similarity values. $\bm{x_{(i,k)}}$ is an $i$th behavioral feature vector of the modified ransomware using $k$th parameter set; we use the 24 combinations of parameters (i.e., 24 rows shown in Fig.~\ref{fig:similarity-mod-mod}). The numerator is a dot product of two vectors, and the denominator is a product of the L2 norm of each vector.

%%%%%%%%%

\begin{figure*}[tb]
\centering
\includegraphics[scale=0.6]{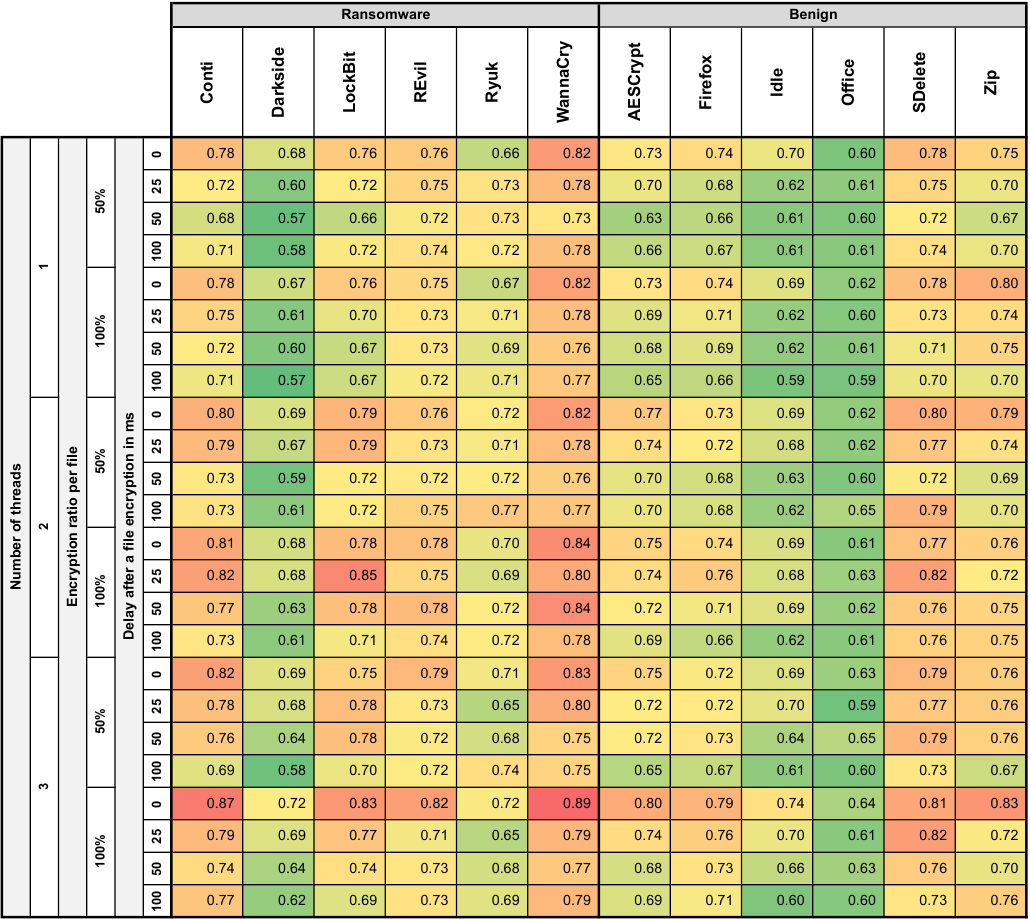}
\caption{Average cosine similarity of 23-dimensional behavioral features between adversarial behavioral features $\chi_{adv}$ of modified Conti ransomware with the micro-behavior control function and 12 samples of the RanSMAP dataset\cite{HIRANO2025104202,RanSMAP-repo}.}
\label{fig:similarity-12-mod}
\end{figure*}

Fig.~\ref{fig:similarity-12-mod} shows the average cosine similarity of all the 23-dimensional behavioral features, $\overline{Sim}_{(k, sample)}$, between modified Conti ransomware with the micro-behavior control function and 12 samples of the RanSMAP dataset\cite{HIRANO2025104202,RanSMAP-repo}.
We calculated the cosine similarity in \eqref{similarity-12-mod}.

\begin{equation}\label{similarity-12-mod}
Sim(\bm{x_{(i,k)}},\bm{x_{(i, sample)}}) 
= \dfrac{\bm{x_{(i,k)}}\,\bm{\cdot}\,\bm{x_{(i,sample)}}}{||\bm{x_{(i,k)}}||\,||\bm{x_{(i,sample)}}||}
\end{equation}

\noindent where $\bm{x_{(i,k)}}$ is an $i$th behavioral feature of the modified ransomware using $k$th parameter set; we use the 24 combinations of parameters (i.e., 24 rows in Fig.~\ref{fig:similarity-12-mod}). $\bm{x_{(i,sample)}}$ is an $i$th behaviraol feature of a sample. The samples comprise six ransomware and six benign applications of the RanSMAP dataset. For example, $\bm{x_{(0,Conti)}}$ is storage entropy feature $\bm{x_0}$ of Conti ransomware. Please note that the Conti ransomware sample of the RanSMAP dataset (shown in the leftmost column) is not the same sample we used in changing ransomware behavior; instead, we used the leaked Conti source code \cite{leaked-conti}. An average cosine simiairy $\overline{Sim}_{(k, sample)}$ was calculated in \eqref{average-similarity}.

\begin{equation}\label{average-similarity}
\overline{Sim}_{(k, sample)}\,=\,\dfrac{1}{23} \sum_{i=0}^{22} Sim(\bm{x_{(i,k)}},\bm{x_{(i,sample)}})
\end{equation}

%%%%%%%%%%%%%%%%%%%%%%%%%%%%%%%%%%%%%%%%%%%%%%%%
%% Success rate of evasion attacks
%%%%%%%%%%%%%%%%%%%%%%%%%%%%%%%%%%%%%%%%%%%%%%%%
\section{Success rate of evasive ransomware attacks}\label{sec:success-rate}

\begin{figure*}[tb]
\begin{center}
\begin{subfigure}[t]{0.32\linewidth} 
\includegraphics[width=1.0\textwidth]{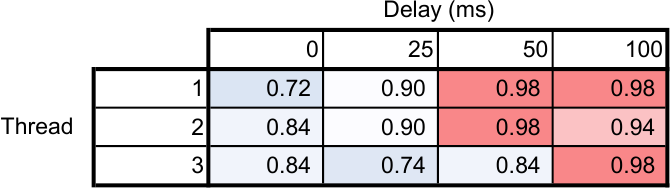}
\centering
\caption{Encryption ratio of 50\%}\smallskip
\label{fig:recall-050}
\end{subfigure}
\begin{subfigure}[t]{0.32\linewidth} 
\includegraphics[width=1.0\textwidth]{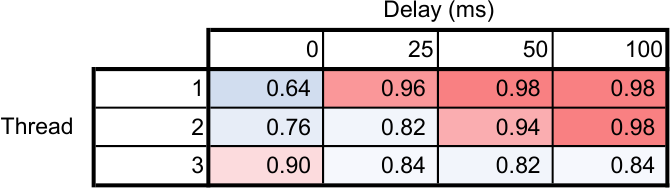}
\centering
\caption{Encryption ratio of 100\%}\smallskip
\label{fig:recall-100}
\end{subfigure}
\end{center}
\caption{Recall of ransomware detector in binary classification (i.e., ransomware or benign) using modified Conti ransomware with the micro-behavior control function. The deep-learning model was trained using the RanSMAP dataset\cite{HIRANO2025104202,RanSMAP-repo}.}
\label{fig:recall}
\end{figure*}

Next, we examined the success rate of evasive ransomware attacks using the micro-behavior control function of modified Conti ransomware.
%% Dataset we used
We trained the deep-learning model using behavioral feature $\bm{\chi}$ from the RanSMAP dataset \cite{HIRANO2025104202,RanSMAP-repo}; we used behavioral features $\bm{\chi}$ of six ransomware (i.e., Conti, Darkside, LockBit, REvil, Ryuk, and WannaCry) and six benign applications (i.e., AESCrypt, Firefox, Idle, Office, SDelete, and Zip) executed on the computer with Intel Core i3 CPU and DDR4 2133Mhz 16GB RAM since we collected adversarial behavioral features $\bm{\chi_{adv}}$ using the micro-control function on the computer with the same specification. We used $T_{window}$\,=\,$0.1$ s and $T_{d}$\,=\,$30$ s to create behavioral feature $\chi$.
%% DL model we used
We created five deep-learning models and tested them for each parameter set of $P_{thread}$, $P_{ratio}$, and $P_{delay}$ and calculated the sum of each confusion matrix.
%% How to evaluate
Recall was used to measure the performance of evasive ransomware attacks since we predicted only positive class (i.e., we input only behavioral features of ransomware with the micro-behavior control function). The recall was calculated in \eqref{eq:recall}.
\begin{equation}\label{eq:recall}
Recall\,=\,\frac{TP}{TP + FN}
\end{equation}
where $TP$ (True Positive) is the number of features in the correct class that are correctly classified. $FN$ (False Negative) is the number of incorrectly classified features that are not in the correct class.
%%
%% Recall and description of the results
Fig.~\ref{fig:recall} shows a recall of modified Conti ransomware with the micro-behavior control function. In Fig.~\ref{fig:recall-050} (i.e., encryption ratio of 50\%), the highest recall of 0.98 was reduced to 0.72 at a minimum; thus, the micro-behavior control function could reduce recall of 0.26. In Fig.~\ref{fig:recall-100} (i.e., encryption ratio of 100\%), the highest recall of 0.98 was reduced to 0.64 at the minimum; thus, the micro-behavior control function could reduce recall of 0.34. 

%%%%%%%%%%%%%%%%%%%%%%%%%%%%%%%%%%%%%%%%%%%%%%%%
%% Discussion
%%%%%%%%%%%%%%%%%%%%%%%%%%%%%%%%%%%%%%%%%%%%%%%%
\section{Discussion}\label{sec:discussion}

We presented the micro-behavior control function to change the behavior of ransomware; this test component simulates source code changes in \eqref{eq:source} to find the optimal source code $\mathcal{S}_{adv}$. 
The experimental results in Fig.~\ref{fig:similarity-mod-mod} show that some features are easy to change and others are difficult. For example, the behavioral features that were easy to change include storage throughput (read) $\bm{x}_2$, MMIO accesses (write) on RAM $\bm{x}_{15}$, MMIO accesses (read) on RAM $\bm{x}_{16}$, variance of GPA on RAM (write) $\bm{x}_{19}$. Please note that we excluded 4KiB page accesses since the number of 4KiB accesses was very small. On the other hand, the behavioral features that were difficult to change include storage entropy $\bm{x}_0$, 2MB page accesses on RAM $\bm{x}_{11}$, $\bm{x}_{12}$, $\bm{x}_{13}$, $\bm{x}_{14}$, and variance of GPA on RAM (instruction fetch) $\bm{x}_{21}$.

Fig.~\ref{fig:similarity-12-mod} shows how an attacker can use behavioral adversarial example $\bm{\chi}_{adv}$ to mimic benign applications. For example, behavioral adversarial example $\bm{\chi}_{adv}$ created using the micro-behavior control function with $\mathit{P_{thread}}\,=\,3$, $\mathit{P_{ratio}}\,=\,100\%$, and $\mathit{P_{delay}}\,=\,25~\mathrm{ms}$ is similar to the behavioral features of SDelete (i.e., benign secure delete program). Please note that each of the 23-dimensional features is not treated equally in deep learning models; therefore, the average cosine similarity is not directly reflected in the result in Fig.~\ref{fig:recall}. The success rate of evasive ransomware attacks can be confirmed in Fig.~\ref{fig:recall}. The recall was reduced to 0.64 at a minimum; however, the success rate is insufficient to complete attacks; further tests using more parameters (e.g., delay of more than 100 ms) are needed.

We discuss the limitations of the presented micro-behavior control function, the test component to simulate changing the source code of evasive ransomware. We formulated how to find the optimal source code $\mathcal{S}_{adv}$ in \eqref{eq:source}; since the search space of possible source code for all possibilities is vast, we developed the micro-behavior control function to add perturbations to behavioral features by changing the number of threads that affect file encryption speed, the encryption ratio that affects entropy values of written data, and the delay after file encryption that affects encryption speed. The three control functions reduced recall but were insufficient for reliable evasive attacks. More flexible and automated behavioral manipulation techniques are needed to cover the ample input space of source code. The fundamental problem to be solved to generate an evasive ransomware's source code is to find reverse mapping from a behavioral adversarial perturbation ${\bm{\epsilon}}$ to a source code $\mathcal{S}$ formulated in \eqref{eq:mapping}; the adversarial perturbation ${\bm{\epsilon}}$ corresponds to a patch (i.e., diff) to create the target source code $\mathcal{S}_{adv}$ from the original source code $\mathcal{S}_{orig}$.

%%%%%%%%%%%%%%%%%%%%%%%%%%%%
\section{Conclusion}\label{sec:conclusion}

Recently, AI-based cybersecurity defense systems have been attacked using adversarial examples; from the defender's perspective, developing a defense system resistant to adversarial examples is crucial. In particular, this paper focused on ``behavioral'' adversarial examples to evade ransomware detectors. We formulated an evasive ransomware attack and examined it using the test component named the micro-behavior control function in Conti ransomware. We tested the evasive attack on the deep-learning-based ransomware detector. The presented attack reduced recall from 0.98 to 0.64; however, the current prototype cannot cover the vast source code space for reliable evasive attacks. Future work includes a more flexible and automated source code generation method that meets the requirements we formulated in this paper; a reliable defense mechanism against the presented evasive attacks will be needed.

%%%%%%%%%%%%%%%%%%%%%%%%%%%%
\section*{Acknowledgment}

This work was supported by JSPS KAKENHI Grant Number JP23H03396 and JSPS KAKENHI Grant Number JP23K11114. The authors gratefully acknowledge constructive comments by the anonymous reviewers. The authors thank Syuya Kawai, Kaito Motokawa, and Reishi Kondo for their support in constructing the dataset. The authors gratefully thank the developers of BitVisor\cite{bitvisor}.

\bibliographystyle{IEEEtran}
\bibliography{hirano}

\end{document}